\documentclass[aps,pre,twocolumn,superscriptaddress,amsmath,amssymb,floatfix,
showpacs]{revtex4}
\usepackage{graphicx}
\usepackage{times}
\begin{document}
\def\d{\delta}
\def\D{\Delta}
\def\s{\sigma}
\def\g{\gamma}
\def\e{\epsilon}
\def\b{\beta}
\def\a{\alpha}

\title{Cliques and duplication-divergence network growth
}
\author{I. Ispolatov}
\email{iispolat@lauca.usach.cl}
\altaffiliation{Permanent address:
Departamento de Fisica, Universidad de Santiago de Chile,
Casilla 302, Correo 2, Santiago, Chile}
\affiliation{Ariadne Genomics Inc,. Rockville, MD 20850}
\author{P. L. Krapivsky}
\email{paulk@bu.edu}
\affiliation{Center for Polymer Studies and
Department of Physics, Boston University, Boston, MA 02215}
\author{I. Mazo}
\email{mazo@ariadnegenomics.com}
\affiliation{Ariadne Genomics Inc,. Rockville, MD 20850}
\author{A. Yuryev}
\email{ayuryev@ariadnegenomics.com}
\affiliation{Ariadne Genomics Inc,. Rockville, MD 20850}

\date{\today}
\begin{abstract}
  A population of complete subgraphs or cliques in a network evolving via
  duplication-divergence is considered. We find that a number of cliques of
  each size scales linearly with the size of the network. We also 
  derive a clique
  population distribution that is in perfect agreement with both the
  simulation results and the clique statistic of the protein-protein binding
  network of the fruit fly.  In addition, we show that such features as
  fat-tail degree distribution, various rates of average degree growth and
  non-averaging, revealed recently for only the particular case of a
  completely asymmetric divergence, are present in a general case of
  arbitrary divergence.

\end {abstract}

\maketitle

\section{introduction}

The duplication-divergence mechanism \cite{japan,bio} of network growth is
traditionally used to model protein networks: A duplication of a node is a
consequence of the duplication of the corresponding gene, and a divergence or
loss of redundant links or functions is a consequence of gene mutations
\cite{pastor,ves,dd,nik}.  General properties of the duplication-divergence
growth 
have recently been studied for probably the simplest version of the
duplication-divergence model which is the asymmetric divergence \cite{us}.
Yet even this simplest model, where links are removed with a certain
probability only from the replica node, turned out to have very rich
phenomenology and to reproduce the degree distribution, observed in real
protein-protein networks, surprisingly well.  Overall, when the link removal
probability is small, the network growth is not self-averaging and an average
vertex degree is increasing algebraically.  For larger values of the link
removal probability, the growth is self-averaging, the average degree
increases very slowly or tends to a constant, and a degree distribution has a
power-law tail.

A natural next step in exploring properties of the duplication-divergence
networks is to consider their modular structure and distribution of various
subgraphs or motifs.  Small subgraphs are often considered building blocks of
network; densities of particular subgraphs may tell if a network belongs to a
certain ''superfamily'' \cite{alon2} or performs specific functions
\cite{alon}. Abundances of triangles and loops have been studied in the
Internet, random and preferential attachment networks and regular scale-free
graphs \cite{triangles,triangles2,internet_loops, reg_loops}.  Densities of
small motifs and cycles centered on a vertex were considered as a function of
the vertex degree and clustering coefficient in \cite{b2}.  In
protein-protein networks, highly interconnected subgraphs were found to be
well-conserved in evolution \cite{barabasi} and to correspond to functional
protein modules in living cells \cite{mirny}. An extreme case of highly
interconnected motifs are cliques, or completely connected subgraphs. Cliques
have been found in higher than random abundances in protein-protein networks
in yeast \cite{mirny}.

In the paper we consider a generalization of duplication-divergence network
growth mechanism, duplication-divergence-heterodimerization.  The
heterodimerization, or linking a certain number of the pairs of target and
replica nodes, is essential for clustering and is observed in protein-protein
networks \cite{paralinks}. We show that
duplication-divergence-heterodimerization produces the cliques in the number
very similar to those observed in protein-protein networks.
 
As in our previous work \cite{us}, we again start with the simplest case of
the completely asymmetric divergence.  Yet in real protein networks, apart
from special cases of partially asymmetric divergence \cite{wagner_asym}, the
divergence is believed to be close to symmetric \cite{sym}.  It turns out
that the asymmetric divergence results for the clique statistics as well as
the previously obtained results for the network growth \cite{us} are
qualitatively similar to those in the arbitrary divergence case, where links
are removed with given probabilities both from the target and replica nodes.

The paper consists of two principle parts: In the next section we derive the
clique abundance distribution for the asymmetric case and compare it to the
simulation and experimental results. In Sec.~III we generalize these and
previously obtained results for the network growth and degree distribution
onto the arbitrary divergence case. A Discussion and Conclusion section
completes the paper.

\section{cliques}

Protein-protein networks exhibit a distinct modular structure and contain
densely linked neighborhoods or complexes (\cite{mirny} and references
therein). The extreme case of densely linked complexes are cliques or
completely connected sub-graphs where each vertex is connected to all other
subset members.  Cliques of the sizes of up to 14 vertices were found in much
higher than ``random'' abundance in protein binding network of yeast
\cite{mirny}.  Yet many large cliques observed in protein networks may be
artifacts of specific experimental techniques or even of misinterpretation of
the experimental data.
%%%%%%%%%%%%%%%%%%%%%%%%%%%%%
For example, there is a strong evidence that all cliques of order higher than
six in the 
yeast interaction network  \cite{badcl} considered in \cite{mirny} result from 
the "matrix" 
recording of the experimental 
data from the mass-spectrometry experiments. 
In such experiments  an immunoprecipitation is used to isolate stable protein
complexes.  Usually a single protein is used as a target for the antibody;
binding of the antibody to this protein leads to an isolation of the 
the entire complex. However, the precise pairwise binding between proteins in
the complexes strictly speaking remains undefined if a complex contains more
than two proteins. Yet in the ''matrix'' interpretation of the
mass-spectrometry experiment 
all possible pairwise 
interactions between proteins in the complex are usually recorded.
A well-known example of such erroneous
recording is the anaphase-promoting
complex.  It is reported as a 
11-node clique in three
different mass-spectrometry 
high-throughput interaction surveys of yeast genome and
in the MIPS database \cite{badcl}.  The biggest reported clique in yeast
network, SAGA/TFIID complex \cite{mirny}, 
is also the result of erroneous "matrix"
recording of the data from a co-immunoprecipitation experiment described in
\cite{saga}.  

However, a virtually free of subjective interference two-hybrid method, 
used
to determine the protein binding network of fruit fly \cite{fly}, yields also
higher than ''random'' number of cliques. Specifically, the fly dataset
contains 1405 triads, 35 4-cliques and one 5-clique, while a randomly
re-wired graph of fly dataset contains only 1147 triads and 8 4-cliques
\cite{rewiring}.  Here and below, the lower-oder cliques that comprise the
higher-order ones (each clique with $j$ vertices or ''$j$-clique'' consists
of $j$ cliques with $j-1$ vertices which can be obtained by eliminating one
of the $j$ vertices) are counted along with the non-trivial cliques. The
number of only non-trivial cliques is slightly lower; the fly dataset
contains 1297 non-trivial triads, 30 4-cliques and one 5-clique.

Is such high concentration of large cliques caused by an evolutionary
pressure that specifically favors big cliques, or by some stochastic
mechanism of network evolution? Evidently, a simple duplication-divergence
network growth never produces even a single triad as new duplicates are never
linked to their ancestors \cite{us}.  Random mutations, or re-wiring of some
links will give rise to a certain number of cliques, yet their abundance will
be much less than the experimentally observed one \cite{mirny, rewiring}.
However, in \cite{paralinks} it was concluded that links between paralogs (or
recently duplicated pairs of proteins) are significantly more common than if
such links appeared by random mutations. Most of these paralogous links are
formed when a self-interacting protein or (homodimer) is duplicated
\cite{paralinks}, thus giving rise to a pair of interacting heterodimers.
While after divergence certain pairs of heterodimers loose their ability to
interact, some paralogs retain their propensity to heterodimerize.
In the following we show that the simple duplication-divergence network
growth complimented with heterodimerization of some pairs of duplicates does
explain the observed abundance of cliques without invoking any evolutionary
pressure.

The duplication-divergence-heterodimerization process consists of
duplication-divergence, previously introduced in \cite{us}, 
\begin{itemize}
\item {\bf Duplication}. A randomly chosen target node is duplicated, that is
  its replica is introduced and connected to each neighbor of the target
  node.
\item {\bf Divergence}. Each link emanating from the replica is activated
  with probability $\sigma$ (this mimics link disappearance during
  divergence). 
\end{itemize}
and heterodimerization,
\begin{itemize}
\item {\bf Heterodimerization}. The target and replica nodes are linked with
  probability $P$. It mimics the probability that the target node is a dimer
 and the propensity for dimerization is preserved during divergence.
\end{itemize}
Similarly to the ''pure'' duplication-divergence growth \cite{us}, the
replica is preserved if at least one link is established; otherwise the
attempt is considered as a failure and the network does not change.

\begin{figure}
\includegraphics[width=.35\textwidth]{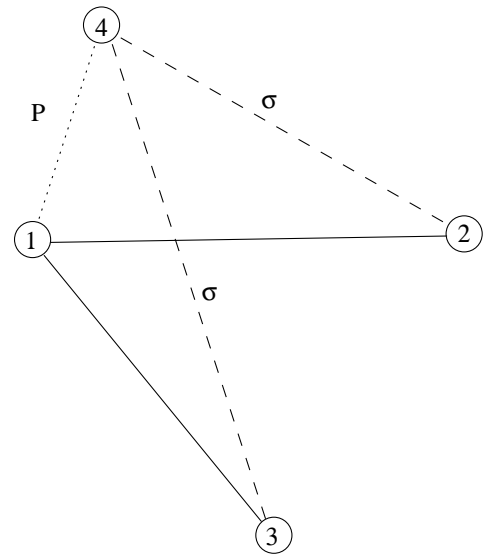}
\caption{\label{triada}
A sketch of duplication event when a new triad is formed with a
heterodimerization link. Solid lines correspond to the existing links, 
dotted line is a heterodimerization link, established with the probability
$P$, and dashed lines denote the inherited with probability $\s$ links.
} 
\end{figure}

\begin{figure}
\includegraphics[width=.35\textwidth]{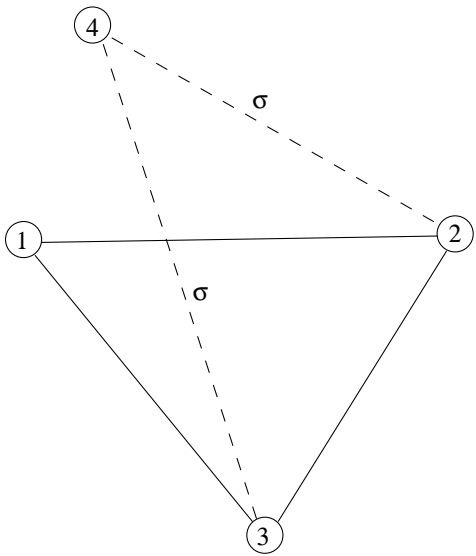}
\caption{\label{triadb}
A sketch of duplication event when a new triad is formed by duplicating the
existing one. Solid lines correspond to the existing links 
and dashed lines denote the links, each inherited with the probability $\s$.
} 
\end{figure}

Let us first consider an evolution of population of triads, or 3-cliques.
Two processes that give rise to new triads are illustrated in Figs.
\ref{triada},\ref{triadb}.  During the first process a target vertex 1,
initially linked to the vertices 2 and 3, is duplicated to produce a new
vertex 4.  The resulting pair of duplicates 1 and 4 have a probability $P$ to
be linked. In addition, links 4-2 and 4-3 are inherited with the probability
$\s$ each. As a result of this process, two new triads 1-4-2 and 1-4-3 are
formed, each with probability $P \s$.
%%%%%%%%%%%%%%%%%%%%%%%%%%%%%%%%%%%%%
%%%%%%%%%%%%%%%%%%%%%%%%%%%%%%%%%%%%
%%%%%%%%%%%%%%%%%%%%%%%%%%%%%%%%%%%%%
%%%%%%%%%%%%%%%%%%%%%%%%%%%%%%%%%%%%
In the second process (Fig.~\ref{triadb}) a new triad is produced from the
existing one when one of its vertices (vertex 1) is duplicated.
The new triad is formed only if both links, 4-2 and 4-3 survive divergence,
which happens with the probability $\s^2$.

Correspondingly, a rate equation for the increase in the number of triads
$C_3$ per duplication-divergence-heterodimerization step contains two terms,
\begin{equation}
\label{tri}
 {\D C_3} =  \s P \frac{ 2 L} {N} + \s^2 \frac{3 C_3} {N};
\end{equation}
where $L$ and $N$ are the numbers of links and vertices in the network.
The fraction $2L/N$ in the first term is an average number of links picked up 
for a potential triad (which is also equal to the average degree $\langle d
\rangle$). The factor 3 in the second term indicates that each of the three
vertices in the existing triad can be picked up as a target vertex for
duplication. 

Considering links as 2-cliques, the first term in Eq.~(\ref{tri})
can be interpreted as describing a creation of 3-clique from a lower-order
2-clique.
It is easy to see that with such interpretation, 
the Eq.~(\ref{tri}) can be generalized to describe the evolution of population
of cliques of an arbitrary order,
\begin{equation}
\label{Cj}
\frac {\D C_j} {\D N} = \frac{ (j-1) C_{j-1} P \s^{j-2}} {\nu N} + 
\frac{j C_j \s^{j-1}} {\nu N}.
\end{equation}
Here $\nu \leq 1$ is an increment in the number of vertices per duplication
step. In the following we focus on a biologically-relevant regime of $0 < \s
<1/2$ where the  average degree $\langle d \rangle$ is constant
or almost constant \cite{us}. In this regime $\nu=2\s$, and
assuming scaling for $C_j$,  $C_j\equiv N c_j$, one obtains a recurrent
relation for the rescaled $j$-clique abundance,
\begin{equation}
\label{cj}
{c_j}=
\frac {(j-1)c_{j-1} \s^{j-3} P} {2 - j \s^{j-2}}.
\end{equation}
For large $j$ the second term in denominator becomes subdominant,
\begin{equation}
\label{cjasympt}
{c_j} \sim
(j-1)!\s^{(j-3)(j-2)/2}\;(P/2)^{j-2}.
\end{equation}
It follows that the relative population of large cliques decays faster than
exponentially. This rends large cliques highly improbable in networks of
biologically relevant size of $N\sim 10^4$.

To check this analytical prediction and to see if the proposed
duplication-divergence-heterodimerization model explains the observed
population of cliques, we performed the following numerical simulation. As in
\cite{us}, we fix $\s=0.38$ so that the average degree is equal to that of
the fly dataset, where $\langle d \rangle \approx 5.9$ for $N=6954$ proteins
\cite{fly}. We select $P=0.03$ so that the number of triads in the simulated
network is also similar to that in the fly dataset and count the number of 4-
and 5-cliques in the resulting network.  The theoretical $C_j$ are computed
for the same $\s$ and $P$ taking into account that $c_2 \equiv \langle d
\rangle/2$.  Results of simulations $C_j^{s}$ averaged over 2000
network realizations, the computed $C_j^{th}$, and the clique abundances in
the fly dataset $C_j^{fly}$ are shown in Table~\ref{tab_cliques}.
%%%%%%%%%%%%%%%% RAW CLIQUES %%%%%%%%%%%%%%%%%%%%%%%%%%%%%
\begin{table}
\begin{ruledtabular}
\begin{tabular}{cccc}
$j$ & $C_j^{fly}$ & $C_j^{s}$ & $C_j^{th}$ \\
\hline
3 &  1405 & $1371 \pm 9$ & 1416 \\
\hline
4 &  35   & $33 \pm 1$   & 34 \\ 
\hline
5 &   1   & $0.37 \pm 0.04$ & 0.34\\
\hline
6 &   0   & 0 & 0.0014\\
\end{tabular}
\end{ruledtabular}
\caption{\label{tab_cliques} 
Number of $j$-cliques in networks with $N=6954$ vertices and 
$L= 20435$ links
for $C_j^{fly}$ -- fruit fly protein-protein
binding  
network, $C_j^{s}$ -- simulations with $\s=0.38$ and $P=0.03$, and  
$C_j^{th}$ -- Eq.~(\ref{cj}) prediction for the same $\s$ and $P$.}
\end{table}
The agreement between the experimental dataset, simulations, and
Eq.~\ref{cj} is surprisingly good, especially given the fact that
in for  $\s=0.38$, $\langle d \rangle =const$ only approximately
\cite{us}.

\section{symmetric divergence}
In this section we generalize the results obtained in \cite{us} and above for
the case of completely asymmetric divergence onto an arbitrary divergence
case.  The arbitrary divergence model is defined as follows:
\begin{enumerate}
\item {\bf Duplication}. A randomly chosen target node is duplicated, that is,
  its replica is introduced and connected to all neighbors of the target
  node.
\item {\bf Divergence}. Each link emanating from either the target or the
  replica node is independently removed with probability $1-\sigma_1$ and
  $1-\sigma_2$, correspondingly.  This mimics disappearance of links during
  divergence from initially indistinguishable target and replica nodes.
  Vertices that lost all their links during this process (this may include
  both the target and the replica vertices as well as their neighbors) are
  discarded.
\end{enumerate}
Unlike in the asymmetric duplication-mutation models, the symmetric growth
model may generate network consisting of more than one disconnected
components.  V\'azquez and co-workers \cite{ves} investigated a symmetric model
which only slightly differs from the fully symmetric version
($\sigma_1=\sigma_2$) of our model.

\subsection{Growth law}

As in \cite{us}, an increment in the number of links $L$ during  a
duplication step is,
\begin{equation}
\label{LN}
\frac{\D L}{\D N}=\frac{2L(\s_1 + \s_2 -1)}{\nu N},
\end{equation}
where $N$ is the number of vertices, $2L/N \equiv \langle k \rangle $ is 
the average number of neighbors or
the average degree, and $0<\nu\leq 1$ is an  
increment in the number of vertices
per step. Assuming that for a large network $\nu$ does not depend on the
network size $N$, we obtain,
\begin{equation}
\label{L(N)}
L(N)\sim N^{2(\s_1+\s_2-1)/\nu}
\end{equation}
As in the asymmetric case, there exist three distinct regimes:
\begin{itemize} 
\item Since at a duplication
step the number of vertices cannot increase by more than one, $\nu\leq 1$ and 
for $\s_1+\s_2 > 3/2$ the growth  of $L(N)$ is superliniear. The average degree
grows as a power-law of a network size, and for sufficiently large networks
the probability to
eliminate all the links and therefore, not to add a vertex at a duplication
step becomes negligible. Hence for large networks $\nu \to 1$ and 
\begin{equation}
\label{L}
L\sim N^{2(\s_1+\s_2-1)}. 
\end{equation}
\item For $\s_1+\s_2\leq 3/4$ and $\s_1 > \s_1 ^*$, $\s_2 > \s_2 ^*$  
(where the lower bounds 
$\s_i^*$ will be determined
  below), 
we observe that the average degree increases logarithmically and
$L \sim N\ln(N)$.  
\item Since 
only linked vertices are counted, the average degree cannot degrease below
unity. Hence even for small link retention probability, $1 < \s_1+\s_2$,
and $\s_1 < \s_1 ^*$, $\s_2 < \s_2 ^*$ the
growth of $L$ is linear, $L\sim N$ and the average degree saturates to a
constant. 
\end{itemize}
\begin{figure}[ht] 
  \vspace*{0.cm} \includegraphics*[width=0.45\textwidth]{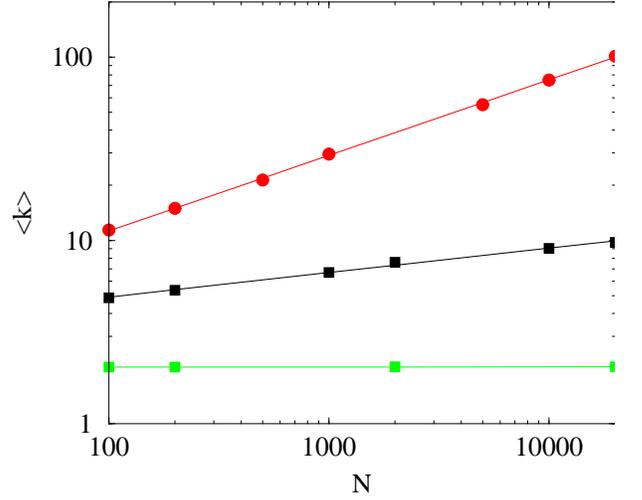}
 \caption{(Color online) The average node degree $\langle k\rangle$ 
   vs $N$ for (bottom to top) for the completely symmetric network growth, 
  $\s_1=\s_2=0.6,0.75,0.85$.  Solid
   lines are corresponding  best
   fits, $\langle k\rangle= const$
   for $\s_1=\s_2=0.6$, $\langle k\rangle \sim N^{0.14}$ or 
   $\langle k\rangle \sim \ln N$ for  $\s_1=\s_2=3/4$,
   and $\langle k\rangle \sim N^{0.41}$ for  $\s_1=\s_2=0.85$
   ($\langle k\rangle \sim N^{0.4}$
   from (\ref{L})). The results
   are averaged over 100 network realizations. }
\label{large}
\end{figure}

%  \begin{figure} 
%   \vspace*{0.cm}
%   \includegraphics*[width=0.45\textwidth]{deg_distr_sym.eps}
%   \caption{(Color online)
%     The degree distribution $n_k$ vs. $k$ for (bottom to to top)
%    $\sigma=0.6$, $\sigma=3/4$, and $\sigma=0.85$. 
%  The size of the network is
%    $N=2 \times 10^4$ and results are averaged over 100 realizations.}
%  \label{degree}
%  \end{figure}

\begin{figure} 
 \vspace*{0.cm}
 \includegraphics*[width=0.45\textwidth]{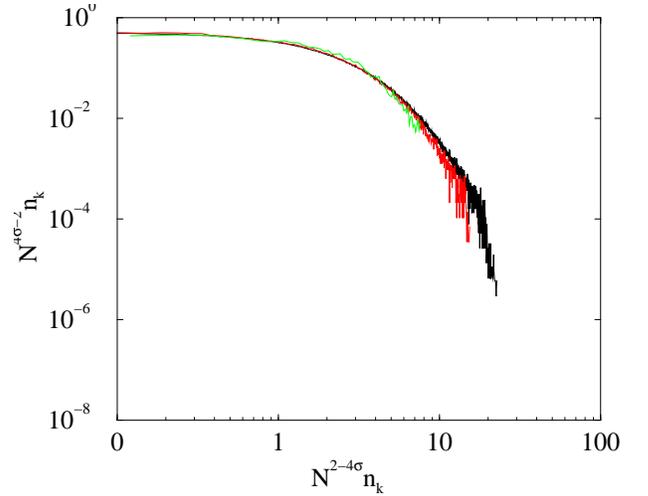}
 \caption{(Color online) Scaling 
   of the degree distribution in the networks of
   $N=200$, $N=2000$, and $N=20000$ nodes with $\s_1=\s_2=0.85$.}
\label{more}
\end{figure}

\subsection{Degree distribution}
\begin{figure} 
 \vspace*{0.cm}
 \includegraphics*[width=0.45\textwidth]{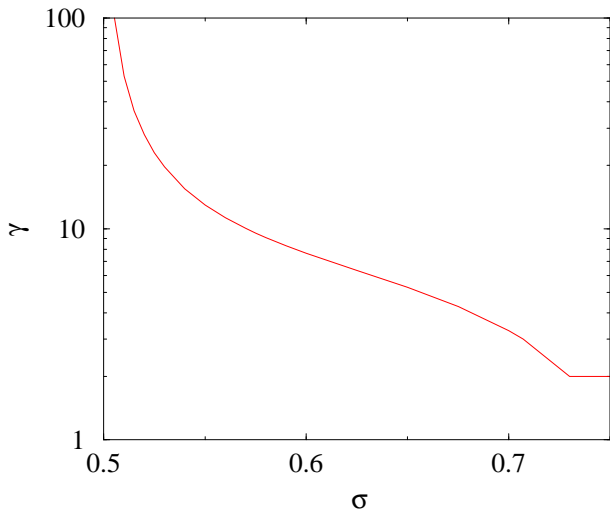}
 \caption{The degree distribution exponent $\g(\s)$ for the symmetric
 divergence from Eq.~(\ref{gamma}), $\gamma \approx 1/(2\s - 1)$
for $\s \to 1/2 +0$}
\label{fig_g}
\end{figure}

\begin{figure} 
 \vspace*{0.cm}
 \includegraphics*[width=0.45\textwidth]{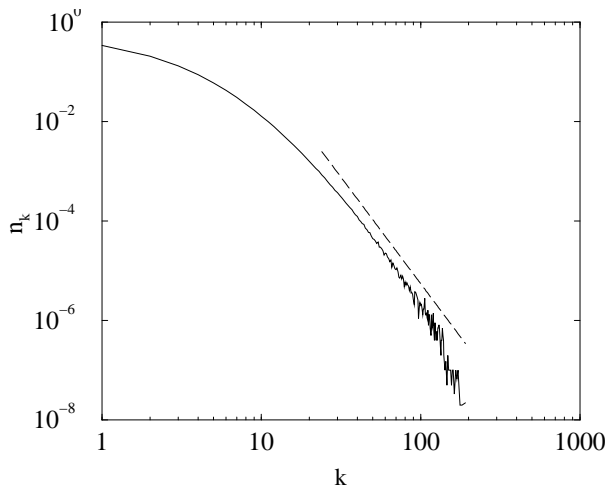}
 \caption{The degree distribution $n_k$ for symmetric divergence, 
$\s_1=\s_2=0.675$. A dashed line is the
 predicted power-law asymptotics with the exponent $\g(0.675)\approx 4.3$.}
\label{degree_distr}
\end{figure}

As in \cite{us}, the degree distribution $N_k$ is described by the following
rate equation, 
\begin{eqnarray}
\label {m}
\nu{\D N_k\over \D N}
\nonumber
={N_{k/\s_1}\over N\s_1} + {N_{k/\s_2}\over N\s_2} -{N_k \over N}\\ 
+(\s_1 + \s_2 -1)\left [{(k-1)N_{k-1}-k N_k\over N} \right ].
\end{eqnarray}
Here the first three terms describe the gain of two new degrees
of the duplicated vertices and the loss of an old degree of the target vertex,
while the fourth term accounts for a change in the number of degrees of
a neighbor of a target vertex.
Substituting $N_k\propto N k^{-\gamma}$ and using $\nu=2(\s_1 + \s_2 - 1)$ 
(which follows
from (\ref{LN})), we obtain
\begin{equation}
\label{gamma}
\s_1^{\g-1}+ \s_2^{\g-1}+ (\s_1 + \s_2 -1)(\g-1) + 1- 2(\s_1+\s_2)=0.
\end{equation}
This equation has a trivial $\gamma'=2$ and a non-trivial solution
$\gamma(\s_1,\s_2)$ 
which intersect at $(\s_1^*,\s_2^*)$ that satisfy 
the equation,
\begin{equation}
\label{sigma}
\s_1( \ln \s_1 + 1) + \s_2( \ln \s_2 + 1)=1. 
\end{equation}
An important example is the symmetric case, $\s_1^*=\s_2^* \approx 0.72985$;
in the asymmetric case $\s_1 \equiv 1$ and $\s_2^*=1/e \approx 0.36879$
\cite{us}. 
The resulting
exponent $\g$ for the symmetric case is plotted in Fig.~\ref{fig_g}.
The measured in simulation 
degree distribution for $1/2<\s<\s^*$ indeed follows the predicted
power-law asymptotics, Fig.~\ref{degree_distr}.

 A summary of  results for the arbitrary-symmetric duplication-divergence
 is presented in Table~\ref{tab_uno}.

\begin{table*}h
\begin{ruledtabular}
\begin{tabular}{cccc}
$\s$ & self-averaging & $L(N)$ & $n_k$ \\
\hline
\hline
$\s_1=\s_2=1$ &  No & $N(N+1)/6$ & $2(N-k)/[N(N-1)]$\\
\hline
$3/2<\s_1 + \s_2 <1$ &  No & $\sim N^{2(\s_1+\s_2-1)}$ & $\sim N^{3-
  2\s_1-2\s_2} 
F(k/N^{2\s_1+2\s_2-3})$\\ 
\hline
$\s_1 + \s_2  < 3/2$, $\s_i > \s_i^*$, $i=1,2$ & Yes & $\sim N \ln N $ &
probably $\sim 
k^{-2}$\\  
\hline
$1/2 < \s_1+  \s_2$, $\s_i < \s_i^*$, $i=1,2$ & Yes & $\sim N $ & $\sim
k^{-\g(\s_1,\s_2)}$\\  
\end{tabular}
\end{ruledtabular}
\caption{\label{tab_uno} 
The behavior of the duplication-divergence network of arbitrary symmetry
for different
values of  
probabilities to preserve a link $\s_1$ and $\s_2$. 
Here $L(N)$ is the average number of links
for given number of nodes $N$, $n_k$ the average fraction of nodes of degree
$k$.
$\s_i^*$, $i=1,2$ are the solutions of Eq.~(\ref{sigma}),  
$\g(\s_1,\s_2) $ is given by Eq.~(\ref{gamma}).}
\end{table*}

\subsection{Cliques}
Similarly to the asymmetric divergence considered above, to generate cliques
one needs to add heterodimerization to the pure duplication and divergence. 
Hence we assume that a target and a replica nodes are linked with
probability $P$.

A generalization of Eq.~(\ref{Cj}) reads
\begin{eqnarray}
\label{Cjg}
\frac {\D C_j} {\D N} = 
\nonumber
\frac{ (j-1) C_{j-1} P (\s_1\s_2)^{j-2}} {\nu N}\\ + 
\frac{j C_j (\s_1\s_2)^{j-1}} {\nu N} - \frac{j C_j (1-\s_1^{j-1})
(1-\s_2^{j-1})} {\nu N}. 
\end{eqnarray}
Since a creation of a new clique requires that all 
links emanating both from the
target and replica vertices survive divergence, in the first two terms
$\s$ is replaced by $\s_1\s_2$. the third term accounts for loss of
$j$-cliques due to disappearance of at least one link both from the
target and replica nodes. Following the procedure  for the asymmetric
case and taking into account that in the scaling regime 
where $1/2 <\s_1+\s_2<3/2$,
$\nu=2(\s_1+\s_2-1)$, we obtain the recurrent relation
(an analog of 
Eq.~(\ref{cj})),
\begin{equation}
\label{cjg}
{c_j}=
\frac {(j-1)c_{j-1} (\s_1\s_2)^{j-2} P} {2(\s_1+\s_2-1) -
  j(\s_1^{j-1}+\s_2^{j-1}-1)}.
\end{equation}

We check this prediction for a completely symmetric case $\s_1=\s_2= \s$, again
using  the fly dataset
\cite{fly} for reference. The correct average degree and number of triads are 
obtained when $\s\approx 0.725$ and $P \approx 0.0475$. The experimental,
simulation, and theoretical results, shown in Table~\ref{tab_cliquessym}, 
are again in very good agreement. 
%%%%%%%%%%%%%%%% RAW CLIQUES %%%%%%%%%%%%%%%%%%%%%%%%%%%%%
\begin{table}h
\begin{ruledtabular}
\begin{tabular}{cccc}
$j$ & $C_j^{fly}$ & $C_j^{s}$ & $C_j^{th}$ \\
\hline
3 &  1405 & $1353 \pm 9$ & 1377 \\
\hline
4 &  35   & $28 \pm 1$   & 28 \\ 
\hline
5 &   1   & $0.24 \pm 0.03$ & 0.24\\
\hline
6 &   0   & $0.0025 \pm 0.0016$ & 0.0011\\
\end{tabular}
\end{ruledtabular}
\caption{\label{tab_cliquessym} 
Number of $j$-cliques in networks with $N=6954$ vertices and 
$L=20435$ links
for $C_j^{fly}$ -- fruit fly protein-protein
binding  
network, $C_j^{s}$ -- simulation of symmetric divergence with
$\s_1=\s_2=0.725$
and $P=0.0475$, and  
$C_j^{th}$ -- Eq.~(\ref{cjg}) prediction for the same $\s$ and $P$.
Simulation results are averaged over 2000 network realizations.}
\end{table}

\subsection{Integrity of the network}

\begin{table}h
\begin{ruledtabular}
\begin{tabular}{ccc}
$\s$ & $n_c$ & $N_L/N$\\
\hline
0.8 &  $ 1.1 \pm 0.01$   & $99 \pm 0.2 \%$ \\
\hline
0.725 &  $ 8.4 \pm 0.2$   & $92 \pm 0.4 \%$ \\
\hline
0.65 &  $232 \pm 1$   & $33 \pm 1 \%$ \\
\hline
0.6  &  $835 \pm 1.4$   & $2.7 \pm 0.03 \%$ \\
\end{tabular}
\end{ruledtabular}
\caption{\label{tab_comp} 
Number of components $n_c$ and the number of vertices in the largest component
normalized by the network size, $N_L/N$, in the duplication-symmetric
divergence networks  
for various $\s_1=\s_2=\s$. All networks are grown to the fly dataset size,
$N=6954$; the results are averaged over 1000 realizations.
}
\end{table}
For symmetric divergence, 
we measure the number of components and the
size of the largest component for the networks grown with various
$\s_1=\s_2=\s$.  The results for the networks of the size of fruit fly
dataset, $N=6954$, are presented in Table~\ref{tab_comp}. 

It follows that for $1/2 < \s \lesssim \s^*$ the grown network consists of
many fairly small components, while for $\s^* < \s$ there is usually one or
few large components and several small ones. Intuitively it is clear that if
the average degree grows, even slowly, the probability to split the network
into many parts becomes small. 

A theoretical prediction for the size of the giant component exists 
only for the Erd\H os-R\'enyi random graph  \cite{jan}:
When the average degree
scales logarithmically with the number of vertices, i.e., $\langle d \rangle=
p \ln N$, the total number of vertices that do not belong to the 
giant component scales as $N^{1-p}$ for $p<1$, while 
for $p>1$ the giant component engulfs the entire system. It turns out
that for the same number of vertices and links, 
the completely random linking of the Erd\H os-R\'enyi graph keeps more
vertices in
a giant component than the corresponding
duplication-symmetric divergence network.
Indeed, for the parameters corresponding to the fly dataset, $\s_1=\s_2=0.725$,
$N=6954$,  $\langle d \rangle \approx 5.9$, and $p=0.667$, the number of
vertices not belonging to the giant component is $6954 \times 0.08 \approx
556$ (see Table~\ref{tab_comp}).  Yet the Erd\H os-R\'enyi graph with the same
number of vertices and links
has only $\sim 6954^{0.333} \approx 19$ vertices outside of its giant
component. This happens mainly because in our duplication-divergence
growth model, once a component is split from the giant component, 
it never re-connects.
If such separation happens at an early stage of the network growth, the
separated component may grow to a significant size, thus leaving
many vertices outside of the giant component. 
On contrary, at each step of the Erd\H os-R\'enyi growth, any two 
components can be united with a random link. This makes the co-existence of
two or more large components very unprobable.
    
\section{discussion and conclusion}
In the previous sections the following conclusions on the clique abundances
and growth laws of 
the duplication-divergence-heterodimerization networks have been made:
\begin{itemize}

\item We showed that the duplication-divergence network
  growth 
model, complimented with heterodimerization links between duplicates,
correctly describes the statistics of cliques in biologically observed 
protein-protein networks. We derive an expression for 
clique population distribution that correctly describes the clique abundances
in the duplication-divergence-heterodimerization networks.
\item Generalizing the results obtained for the completely asymmetric
  divergence 
  in \cite{us}, we demonstrated that similar regimes, such as
  presence and lack
  of self-averaging, growth and saturation of the average degree, scaling and
  fat tail in the degree distribution,  exist in general
  duplication-divergence case as well. In addition, a clique density
  distribution is generalized onto the arbitrary divergence scenario.
\end{itemize}
The heterodimerization links are not taken into account in our description of
the network growth and degree distribution. Despite their crucial role in the
network topology and clique formation, they constitute only about 1\% of all
links and do not contribute significantly to the degrees of the most of the
vertices.  For link inheritance and heterodimerization probabilities $\s$ and
$P$, corresponding to the fly dataset, the resulting number of heterodimeric
links in a network of the size of the fly dataset is $L_{hd}\approx P N/(2\s)
\approx 270$. This is somewhat higher than the observed number of links
between the pairs of recently duplicated (paralogous) proteins
$L_{hd}^{fly}=142$ \cite{paralinks}.  The main reason for this discrepancy is
that in our simulation all heterodimeric links are counted, while in the real
protein network one can reliably identify only the pairs of recently
duplicated proteins.

In a case of not completely asymmetric divergence when links can disappear
both from the target and replica nodes, a network may fragment into several
components. Yet the biological protein networks are believed to be connected
to ensure their functionality. Hence during {\it in vivo} 
divergence the steps that lead to
breaking the network into isolated components are excluded due to evolutionary
pressure. Our probabilistic network growth model does not take any
evolutionary pressure into account. 
However, since for sufficiently high link retention probabilities the resulting
network consists of one or very few large components, the number of
link eliminations that have to be evolutionally overridden is small.
Hence most of the properties of the probabilistically 
grown graphs should be similar to those
of the realistic evolutionary single-component networks.
As the link inheritance probabilities $\s_i$ 
decrease and the number of network
components grow, the number of link removal steps that have to be
evolutionally overridden becomes large. 
Consequently, the probabilistic multi-component
network becomes less similar to the real single-components one.   

As we mentioned in the Section II, we selected the fly dataset as an example
as being the most non-subjective one.
Other know protein-protein networks,
such as for Yeast, Worm, and Human, do contain parts of data that are results
of the "matrix" recording of the experimental 
data from the immunoprecipitation experiments. These datasets contain a higher 
number of large cliques which can be attributed to this data interpretation.
In principle, the clique population distribution derived here can be used to
verify and filter the experimental datasets, removing the erroneously
recordered large cliques.

In a recent publication, Middendorf et al \cite{wiggins} compared topological
properties of the fly dataset to those of the networks grown by 
several mechanisms
such as different versions
of duplication-mutation model and preferential attachment.
It was found that a duplication-mutation-complementation network provides the
best fit to the fly dataset. The duplication-mutation-complementation
network growth model is very close to the
duplication-divergence-heterodimerization model studies here. Complementation
is equivalent to heterodimerization, the only
difference between two models is in the way the links are deleted during
divergence (or mutation): 
Unlike our model,
in \cite{wiggins} each neighbor remains connected to at least 
one of the two duplicates. Thus we confirmed the conclusions made in
\cite{wiggins} that practically all considered properties of protein-protein
networks are very well described by the
duplication-divergence-heterodimerization model.

And finally a few words on the importance of heterodimerization links in
clique formation. An alternative to heterodimerization way to connect
paralogs is to link them randomly by ''mutation'' links. In this case the
probability to establish a heterodimeric link $P$ has to be replaced by a
probability that a mutation link, emanating from a target node, selects the
replica node out of $N$ network nodes. This probability is equal to $M/N$
where $M$ is the number of mutation links established at each duplication
step. In the example of the fruit fly dataset where $P=0.03$ and $N=6954$,
one needs $M=NP=209$ random links at each step to form the correct number of
triads and higher cliques. Obviously, the mutation scenario which requires so
many additional links is completely ruled out due to, for example, average
degree constraint.

\section{acknowledgment}
We are thankful to S.~Maslov for interesting discussions and bringing the
reference \cite{wiggins} to our attention.
This work was supported by 1 R01 GM068954-01 grant from NIGMS.

\end{document}